\documentstyle[aps,prl,multicol]{revtex} 
\input{epsf}
\begin{document}

\draft \title{Ordered and self--disordered dynamics of holes and
  defects in the one--dimensional complex Ginzburg--Landau equation}

\author{Martin van Hecke$^1$ and Martin Howard$^{2}$}
\address{$^1$Center for Chaos and Turbulence Studies, The Niels Bohr
  Institute, Blegdamsvej 17, 2100 Copenhagen \O, Denmark}
\address{$^2$Department of Physics, Simon Fraser University, Burnaby,
  British Columbia, Canada V5A 1S6}

\date{\today} \maketitle
\begin{abstract} 
  We study the dynamics of holes and defects in the 1D complex
  Ginzburg--Landau equation in ordered and chaotic cases. Ordered
  hole--defect dynamics occurs when an unstable hole invades a plane
  wave state and periodically nucleates defects from which new holes
  are born. The results of a detailed numerical study of these
  periodic states are incorporated into a simple analytic description
  of isolated ``edge'' holes. Extending this description, we obtain a
  minimal model for general hole--defect dynamics. We show that
  interactions between the holes and a self--disordered background are
  essential for the occurrence of spatiotemporal chaos in hole--defect
  states.
\end{abstract}


\begin{multicols}{2}  
\narrowtext

\begin{figure} \vspace{-0.2cm}
  \epsfxsize=1.003\hsize \mbox{\hspace*{-.07 \hsize}
    \epsffile{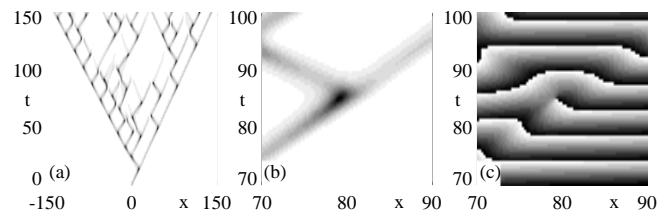} } \vspace{-3.1cm}
\caption[]{(a) A space--time grey--scale plot of $|A|$ (dark:
  $A\!\approx\!0$), 
  showing the propagation of incoherent holes into a plane wave state.
  The dark dots correspond to defects. Note the roughly constant
  velocities at which the holes propagate. Parameter values are
  $c_1\!=\!0.6, c_3\!=\!1.4$ , with an initial condition given by
  Eq.~(\ref{peakeq}), with $\gamma\!=\!1, q_{ex}\!=\!-0.03$.  This
  non--zero $q_{ex}$ breaks the left--right symmetry and results in
  the differing periods of the left and right moving edge holes.
  (b--c) Close--up of $|A|$ (b) and the complex phase $\psi$ (c).
  }\label{fig1}
\end{figure}

The formation of local structures and the occurrence of spatiotemporal
chaos are the most striking features of pattern forming systems. The
complex Ginzburg--Landau equation (CGLE)
\begin{equation} 
  A_t = A + (1+ ic_1) \nabla^2 A - (1-i c_3) |A|^2 A \label{cgle}
\end{equation} 
provides a particularly rich example of these phenomena. The CGLE
describes pattern formation near a Hopf bifurcation and has become a
paradigmatic model for the study of spatiotemporal chaos
\cite{CH,chao1,2d,3d,phasediagram,mvh,maw}.  {\em Defects} occur when
$A$ goes through zero and the complex phase $\psi\!:=\!\arg(A)$ is no
longer defined. In two and higher dimensions, such defects can only
disappear via collisions with other defects, and act as long--living
seeds for local structures like spirals \cite{2d} and scroll waves
\cite{3d} whose instabilities lead to various chaotic states
\cite{2d,3d}.  For the 1D CGLE, however, defects occur only at
isolated points in space--time (see Fig.~\ref{fig1}) and intricate
dynamics of defects and local {\em hole} structures occurs, especially
in the so--called intermittent and bi--chaotic regimes
\cite{phasediagram}. The holes are characterized by a local
concentration of phase--gradient $q\!:=\partial_x \psi$ and a
depression of $|A|$ (hence the name ``hole''), and dynamically connect
the defects (Fig.~\ref{fig1}).  We divide these holes into two
categories: {\em coherent} and {\em incoherent} structures.

{\em Coherent structures -} By this we mean uniformly propagating
structures of the form $A(x,t)= e^{- i \omega t} \bar{A}(x-vt)$
\cite{saar}. Recently, hole solutions of this form called {\em
  homoclinic holes} were obtained \cite{mvh}. Asymptotically,
homoclinic holes connect identical plane waves where $A\!\sim \!e^{i(
  q_{ex} x -\omega t)}$. With $c_1, c_3$ and $q_{ex}$ fixed,
unique left moving and unique right moving coherent holes are
found. Left (right) moving holes with $q_{ex}=Q$ ($q_{ex}=-Q$) are
related by the left--right $ q\! \leftrightarrow \!-q$ symmetry of
the CGLE. Coherent holes have {\em one} unstable core
mode \cite{mvh}.

\begin{figure} \vspace{-.20cm}
  \epsfxsize=.97\hsize \mbox{\hspace*{-.05 \hsize} \epsffile{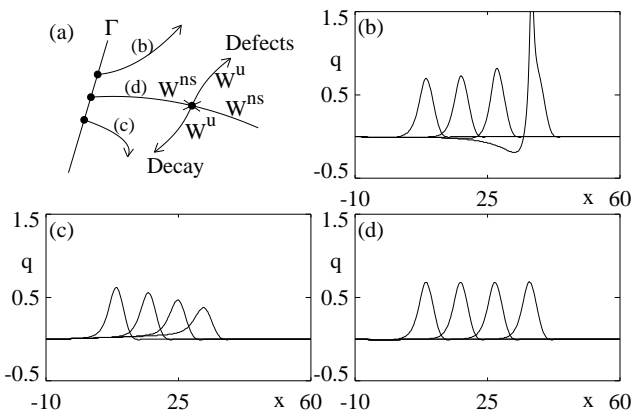}
    } \vspace{-0.4cm}
\caption[]{(a) Schematic representation of the phase space of the CGLE
  around the homoclinic hole solution, showing: the 1D unstable
  manifold $W^{\rm u}$; the high dimensional neutral/stable manifold
  $W^{\rm ns}$ that separates decaying from defect forming states; the 
  manifold $\Gamma$ representing the family of peaked initial
  conditions of the form (\ref{peakeq}).  (b--d) Four snapshots
  $(\Delta t \!=\!10)$ of the $q$-profile of a right moving hole where
  $q_{ex}\!=\!0$, $c_1\!=\!0.6$, and $c_3\!=\!1.4$.  The peaked
  initial condition is given by Eq.~(\ref{peakeq}): (b) A hole
  evolving to a defect $(\gamma=0.568)$, (c) a decaying hole
  $(\gamma=0.5)$, and (d) a hole evolving close to a coherent
  structure $(\gamma=0.5545)$.  }\label{fig2}
\end{figure}

{\em Incoherent structures -} In full dynamic states of the CGLE, one
does not observe the unstable {\em coherent} homoclinic holes, unless
one fine--tunes the initial conditions (see Fig.~\ref{fig2}d). Instead
evolving {\em incoherent} holes that can 

\vspace{0mm}

\noindent grow out to defects occur (Fig.~\ref{fig1} and \ref{fig2}b).

In this Letter we study the hole $\!\rightarrow\!$ defect and defect
$\!\rightarrow\!$ holes dynamical processes of the 1D CGLE
\cite{long}.  We present a minimal model for hole--defect dynamics
that describes the full ``interior'' spatiotemporal chaotic states of
Fig.~\ref{fig1}a, where holes propagate into a self--disordered
background.  Similar ``self--replicating'' patterns are observed in
many other situations, e.g., reaction--diffusion models \cite{rd},
film--drag \cite{gollub}, eutectic growth \cite{faivre}, forced
CGLE \cite{chatepikovsky} and space--time intermittency models
\cite{oldchate}.

{\em Hole $\!\rightarrow\!$ defect -} Let us consider the short--time
evolution of an isolated hole propagating into a plane wave state.
Holes can be seeded from initial conditions like:
\begin{equation}
A=\exp(i[q_{ex}x+(\pi/2)\tanh(\gamma x)])~.\label{peakeq}
\end{equation}
The precise form of the initial condition is not important here as
long as we have a one--parameter family of localized phase--gradient
peaks. This is because the left moving and right moving coherent holes 
for fixed $c_1,c_3$ and $q_{ex}$ are each unique and have one unstable 
mode only. As $\gamma$ is varied three possibilities can arise for the 
time

\vspace{-0.4cm}
\noindent evolution of the initial peak: evolution towards a defect 
(as in Fig.~\ref{fig1}a), decay, or evolution arbitrary close to a 
coherent homoclinic hole (see Fig.~\ref{fig2}). 

The hole propagation velocities
are much larger than the typical group velocities in the plane wave
states: the holes are thus only sensitive to the 
leading wave. Their
internal, slow dynamics determines their trailing wave. A (nearly)
coherent hole will, due to phase conservation, have a trailing wave
(nearly) identical to the leading wave (Fig.~\ref{fig2}); hence the
relevance of the homoclinic holes.

{\em Defect $\!\rightarrow\!$ holes -} What dynamics occurs after a
defect has been formed? A study of
the spatial defect profiles reveals that they consist of a negative
and positive phase--gradient peak in close proximity (the early stage
of the formation of these two peaks can be seen in Fig.~\ref{fig2}b;
see also Fig.~4d of \cite{mvh}). The negative (positive) phase
gradient peak generates a left (right) moving hole. The lifetimes of
these holes depend on their parent defect profile (analogous to what
we described in Fig.~\ref{fig2}) and also on $c_1,c_3$ and
$q_{ex}$. Hence the defects act as seeds for the generation of
daughter holes (see also Fig.~\ref{fig1}). 

\begin{figure} 
 \epsfxsize=1.1\hsize \mbox{\hspace*{-.05 \hsize} \epsffile{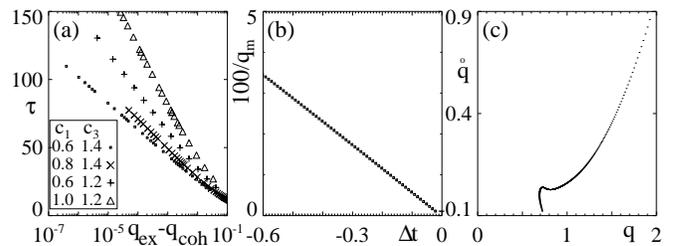} }
 \vspace{-4.0cm}\hspace{-1.cm}
\caption[]{(a)  Log--linear plot of the period $\tau$ as a
  function of $q_{ex}-q_{coh}$.  (b) $100/q_m$ as a function of the
  time $\Delta t$ before the formation of a defect; (c) $\dot{q}_m $
  as a function of $q_m$.  }\label{fig3}
\end{figure}

{\em Periodic hole-defect states -} When an incoherent hole invades a
plane wave state and generates defects, stable periodic hole
$\!\rightarrow\!$ defect $\!\rightarrow\!$ hole behavior can set in
at the edges of the resulting pattern \cite{note2}
(Fig~\ref{fig1}a). 
The asymptotic period $\tau$ of this process depends on $c_1, c_3$,
the propagation direction and the wavenumber $q_{ex}$ of the initial
condition only; we focus here on right moving holes.
%
The period $\tau$ diverges at a well--defined value of
$q_{ex}\!=\!q_{coh}$ (Fig.~\ref{fig3}a). This can be understood in the
phase space picture presented in Fig.~\ref{fig2}.  Suppose we fix
$c_1$ and $c_3$.  The edge defects that are generated periodically
yield 

\vspace{-4mm}
\noindent 
constant initial conditions for their daughter edge holes, similar to
fixing $\gamma$ in Eq.~(\ref{peakeq}).  The period $\tau$ will depend
on the location of the defect profile with respect to the stable
manifold of the coherent hole. When $q_{ex}$ is varied, both this
manifold and the defect profile may change, and for a certain value of
$q_{ex}$ which we call $q_{coh}$, the defect generates an initial
condition precisely on the stable manifold of the {\em coherent
hole}. The lifetime of the resulting daughter hole then diverges (see
Fig.~\ref{fig2}d). 

To substantiate this intuitive picture, we have performed numerics on
the dynamics of ``edge--holes'' invading a plane wave state where $A
\sim e^{i(q_{ex}x -\omega t)}$.  We have performed runs for many
different parameters, but will only discuss a representative subset
here.  Our results indicate that the $\tau$ divergence is of the form
\begin{equation} \label{logdiv}
  \tau \sim -s \ln(q_{ex}-q_{coh})+ \tau_0~.
\end{equation}
This equation, and in particular the value of $s$ can be understood by
considering the flow near the saddle point shown in Fig~\ref{fig2}a.
Just after the hole has been formed, it first evolves rapidly along
the stable manifold. Secondly it evolves slowly along the unstable
manifold before being shot away towards the next defect. For values of
$q_{ex}$ close to $q_{coh}$, the holes approach the coherent structure
fixed point very closely, and $\tau$ will be dominated by a regime of
exponential growth close to this fixed point. Small changes in
$q_{ex}$ will have a negligible effect on the duration of the first
phase ($\tau_0$), but the duration of the second phase will diverge
logarithmically as $~-(1/\lambda) \ln(q_{ex}-q_{coh})$.  Here
$\lambda$, which depends on $c_1$ and $c_3$, denotes the unstable
eigenvalue of the coherent structures at $q_{ex}\!=\!q_{coh}$.  In
Table 1 we list some numerically determined values for $q_{coh}$,
$1/\lambda$, and $s$.  We obtained $s$ and $q_{coh}$ from a fit of
$\tau$ to Eq.~(\ref{logdiv}), whereas $\lambda$ is obtained from a
shooting algorithm, see Ref.~\cite{mvh}.  The agreement between $s$
and $1/\lambda$ is quite satisfactory.

We will now construct a phenomenological model for isolated incoherent
holes.  {\em{(i)}} We will ignore their early time attraction to the
unstable manifold, and think of their location on $W^{\rm U}$ as an
internal degree of freedom, parameterized by the phase--gradient
extremum $q_m$. {\em{(ii)}} Clearly the model should have an unstable
fixed point for values of $q_m$ corresponding to coherent holes. We
have found that, in good approximation, coherent holes have

\begin{figure}
\begin{center}
\begin{tabular}{|c|c|c|c|c|}
  \hline
  ~$c1$~~ & ~$c3$~~ & ~$q_{coh}$ & ~1/$\lambda$~~ &~$s$~~ \\
  \hline
  ~0.6~& ~1.4~& ~-0.0362 ~  &~ 8.42~  &     ~ 8.4~\\
  ~0.8~& ~1.4~& ~-0.0727 ~  &~ 9.91~  &     ~ 9.5~\\
  ~0.6~& ~1.2~& ~0.0538  ~ & ~12.72~ &      ~12.7~ \\
  ~1.0~& ~1.2~& ~-0.0200~ &  ~17.71~&       ~18.7~\\
  \hline
\end{tabular}
\end{center}
Table 1. {\small Comparison of $1/\lambda$ with $s$ (see text for
details).} 
\end{figure}

\noindent 
$q_m\!=\!  q_n\! +\! g q_{ex}$ where $q_n$ denotes the value of $q_m$
for a coherent hole in a $q_{ex}\!=\!0$ state, and $g$ is a negative
phenomenological constant. {\em{(iii)}} When approaching a defect,
$q_m$ diverges as $(\Delta t)^{-1}$ \cite{1/t}; we have confirmed this
by accurate numerics
(Fig.~{\ref{fig3}b}). An appropriate
equation incorporating these three features is

\begin{equation}
  \dot q_m=\lambda (q_m-(q_{n}+ g q_{ex})) +\mu(q_m-(q_{n}+ g
  q_{ex}))^2 ~ ,
  \label{qdot}
\end{equation}
where $g$ and $\mu$ are phenomenological constants.  The first term on
the RHS of (\ref{qdot}) results from the linearization near the
coherent fixed point. Nonlinear terms of higher than quadratic order
on the RHS of Eq.~(\ref{qdot}) are ruled out by the $(\Delta t)^{-1}$
divergence of $q_m$. 
Our numerical data for $\dot q_m$ versus $q_m$ indeed shows quadratic
behavior for large enough values of $q_m$ (Fig.~\ref{fig3}c).
For smaller values of
$q_m$, the curves are quite intricate; this corresponds to the rapid
early time evolution along the stable manifold not included in model
(\ref{qdot}). From Eq.~(\ref{qdot}), it is straightforward to show
that the hole lifetime $\tau$ (the time taken for $q_m$ to diverge)
displays the required logarithmic divergence as $q_{ex}$ is tuned
towards a critical value $q_{coh}$.

{\em Disordered dynamics - } If the patches away from the
holes/defects were simply plane waves with fixed wavenumber, then one
would expect, following the arguments given above, quite regular
dynamics. The coupling between holes and the background induced by
phase conservation becomes the key ingredient to understand disorder
in hole--defect dynamics such as shown in Fig~\ref{fig1}a. Let us
introduce a variable $\phi := \int dx q$ that measures the
phasedifference across a certain interval.

Consider again an edge hole evolving towards a defect. While the peak
of the $q$-profile grows, the hole creates a dip in its wake (see
Fig.~\ref{fig2}b) in order to locally conserve $\phi$. Clearly the
trailing edge of this incoherent hole is {\em not} a {\em perfect}
plane wave. In the interior of states such as shown in
Fig.~\ref{fig1}a, unstable holes move back and forth through a
background of disordered $q_{ex} $ and amplify this disorder.
Nevertheless, as we pointed out earlier, the disordering dynamics is
sufficiently slow such that the holes remain approximately homoclinic
for much of their lives. Although the typical range of values for the
disordered $q_{ex}$ is small, the hole lifetimes depend on it
sensitively. Hence the variation in $q_{ex}$ and $\phi$ is sufficient
to explain the varying lifetimes found in the interior states such as
that shown in Fig~\ref{fig1}a. Thus the essence of the spatiotemporal
chaotic states here lies in the {\em propagation of unstable local
  structures in a self--disordered background}.
   
{\em Minimal model -} To illustrate our picture of self--disordered
dynamics, we will now combine the various hole--defect properties
with the left--right symmetry and local phase conservation of the
CGLE to form a minimal model of hole--defect dynamics. From our
previous analysis, we see that the following hole--defect properties
should be incorporated: {\em{(i)}} Incoherent holes propagate either
left or right with essentially constant velocity (see
Fig.~\ref{fig1}a). {\em{(ii)}} For fixed $c_1,c_3$, their
lifetime depends on the profile of their parent defect, the direction
of propagation, and on the
wavenumber of the state into which they propagate. {\em{(iii)}}
Eq.~(\ref{qdot}) captures essentially all aspects of the evolution of
their internal degree of freedom. When $q_m$ diverges, a defect
occurs. 

In our model we will assume that all the defects have the same profile
and so act as unique initial conditions for their daughter incoherent
holes. While in principle a defect profile could depend on the entire
history of the hole which preceded it, for simplicity we have chosen
to neglect this. We have observed that for some regions of the $c_1,
c_3$ parameter space, the defect profiles from the interior
spatiotemporal chaotic patterns show a surprising lack of scatter
\cite{long}. Therefore we believe that treating the defect profiles as
constant, and only including the effect of the background in the hole
dynamics incorporates the essence of the coupling to a disordered
background.

\begin{figure} \vspace{.cm}
 \epsfxsize=1.\hsize  
 \mbox{\hspace*{-.05 \hsize} \epsffile{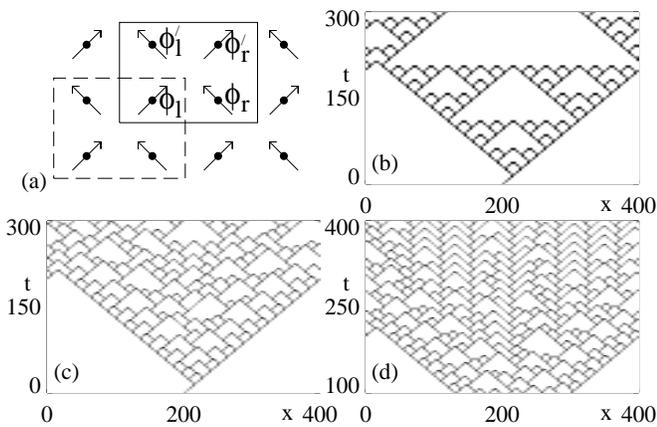}
   } \vspace{-.5cm}
\caption[]{(a) Grid model geometry showing the sites (dots) and 
  hole propagation direction (arrows). The update rule is defined
  within a $2\times 2$ cell. (b--d) Dynamical states in the grid
  model, for $\phi_n\!=\!0.6$ and $\phi_{ad}\!=\!0.75$. Initial
  condition: center site has $\phi\!=\!0.7 $, everywhere else
  $\phi\!=\!0$ (hence the symmetric patterns).  (b) $g\!=\!0$ and
  $\phi_d\!=\!1$. (c) Disordered dynamics for nonzero coupling
  ($g\!=\!-3, \phi_d\!=\!1$). (d) Zigzag structures occur for
  $g\!=\!-3,\phi_d\!=\!0.98$.  }
\label{fig4}
\end{figure}

We discretize both space and time by coarse-graining, and take a
``staggered'' type of update rule which is completely specified by the
dynamics of a $2 \times 2$ cell (see Fig.~\ref{fig4}a). We put a
single variable $\phi_i$ on each site, corresponding to the phase
difference across a cell divided by $2 \pi$.  Local phase conservation
is implemented by $\phi'_l\!+\!\phi'_r \!=\!\phi_l\!+ \!\phi_r$, where
the primed (unprimed) variables refer to values after (before) an
update.

Holes are represented by active sites where $|\phi|>\phi_t$; here
$\phi$ plays the role of the internal degree of freedom. Inactive
sites are those with $|\phi|<\phi_t$, and they represent the
background. The value of the cutoff $\phi_t$ is not very important as
long as it is much smaller than typical values of $\phi$ for coherent
holes. Here $\phi_t$ is fixed at $0.15$. Without loss of generality we 
force holes with positive (negative) $\phi$ to propagate only from
$\phi_l$ ($\phi_r$) to $\phi'_r$ ($\phi'_l$).

Depending on the two incoming states, we have the following three
possibilities:

{\em One site active: } Without loss of generality we assume that we
have a right moving hole.  We implement evolution similar to
Eq.~(\ref{qdot}), but neglect the quadratic term of Eq.~(\ref{qdot});
even though $q_m$ diverges, the local phasedifference $\phi_m$ does not
diverge near a defect. Hence the finite time divergence of the local
phase gradient $q$ that signals a defect can be replaced by a cutoff
$\phi_d$ for $\phi$. Therefore, when $\phi_l \!<\!  \phi_d$, the
internal hole coordinate $\phi$ is taken to evolve via
$\phi'_r\!=\phi_l+\!  \lambda(\phi_l-\phi_n-g\phi_r)$. Here $\lambda$
sets the time scales and can be taken small (fixed at $0.1$). This
evolution equation, combined with the local phase conservation,
means that an incoherent hole propagating 

\noindent into a perfect
laminar state will leave a disordered state in its wake.  When $\phi_l
> \phi_d$, a defect occurs and two new holes are generated:
$\phi'_r=\phi_{ad}$, and $\phi'_l= \phi_d-1-\phi_{ad}$. The factor
$-1$ reflects the change in winding number at a defect.

{\em Both sites inactive: } Away from the holes/defects, the relevant
dynamics is phase diffusion. This is implemented via: $\phi'_r \!=\!
D \phi_l \!+\! (1\!-\!D) \phi_r$. The value of $D$ is fixed at $0.05$
and is not very important.

{\em Both sites active: } This corresponds to the collision of two
oppositely moving holes. Typically this leads to the annihilation of
both holes (see Fig.~\ref{fig1}a), which we implement here via phase
conservation: $\phi'_r \!=\!  \phi'_l \!=\! (\phi_l \!+\!  \phi_r)/2$.

The coupling of the holes to their background, $g$, should be taken
negative (although its precise value is unimportant). For $g\!=\!0$
the lifetime $\tau$ becomes a constant, independent of the $\phi$ of
the state into which the holes propagate, and moreover, the dynamical
states are regular Sierpinsky gaskets
(Fig.~\ref{fig4}b). Nevertheless, starting from a $\phi\!=\!0$ state,
the local phase conservation of the hole dynamics leads to a
background state with a disordered $\phi$ profile.  For $g\!<\!0$
the coupling to this background leads to disorder as shown in
Fig.~\ref{fig4}c,d. This illustrates the crucial importance of the
coupling between the holes and the self--disordered background.

The essential parameters determining the qualitative nature of the
overall state are $\phi_n$, $\phi_d$ and $\phi_{ad}$. These parameters
determine the amount of phase winding in the core of the
$q_{ex}\!=\!0$ coherent holes ($\phi_n$) and in the new holes
generated by defects ($\phi_{ad},\phi_d-1-\phi_{ad}$). When varying
the CGLE coefficients $c_1,c_3$, these parameters change too; for
example, $\phi_n$ typically decreases when $c_1$ or $c_3$ are
increased.  As a result, for large values of $c_1$ and $c_3$,
$|\phi'_l|$ and $\phi'_r$ are typically larger than $\phi_n$ so that
most ``daughter holes'' will grow out to form defects and hole-defect
chaos spreads (Fig.~4c,d). For sufficiently small values of $c_1$ and
$c_3$, on the other hand, $\phi_n$ is large and both daughter holes
will decay. For intermediate values of $c_1$ and $c_3$ it may occur
that $|\phi'_l|$ is significantly larger than $\phi'_r$, leading to
zigzag states [6] (Fig.~4d).


In conclusion, we have studied in detail the dynamics of local
structures in the 1D CGLE. We have obtained a quantitative
understanding of the edge holes, unraveled the interplay between
defects and holes, and put forward a simple model for some of the
spatiotemporal chaotic states occurring in the CGLE.

M.v.H. acknowledges support from the EU under contract 
ERBFMBICT 972554.
M.H. acknowledges support from the Niels
Bohr Institute, the NSF through the
Division of Materials Research, and NSERC of Canada.

\end{multicols}


\begin{references}
\bibitem{note2} The full state that develops in the wake of the
  incoherent holes is chaotic, i.e. one has exponential sensitivity.
  For the edge holes, however, local perturbations of the initial
  condition lead only to a space--time shift, and the asymptotic
  period $\tau$ is independent of such perturbations.
\bibitem{1/t} To see this note that $A$ is a smooth function of $x$
  and $t$, and express $q$ as a function of the real $(u)$ and
  imaginary $(v)$ parts of $A$: $q\!=\! (u\partial_x v- v\partial_x u
  )/|A|^2$. As a function of time, $u$ and $v$ are both moving
  linearly through zero at the defect, and hence we find a $1/\Delta
  t$ divergence of $q_m$.
\end{references}
\end{document}